\documentclass[aps,prb,onecolumn,preprint,superscriptaddress]{revtex4}

\usepackage[dvips]{graphics}
\usepackage{psfig}

\begin{document}

\title{ Quasiparticle Trapping In Three Terminal Ferromagnetic Tunneling Devices }

\author{R. Latempa}
\email{rossella.latempa@mdm.infm.it}
\affiliation{INFM, Departamento di Fisica, Universit\`{a} di Salerno, via S.
Allende, 84081 Baronissi (SA), Italy. }

\author{M. Aprili}
\affiliation{Laboratoire de Physique des Solides, UMR 8502, Universit\'{e} Paris-Sud,
B\^{a}t. 510, 91405 Orsay Cedex, France.}
\affiliation{Laboratoire de Physique Quantique, ESPCI, 10 rue Vauquelin, 75005
Paris, France}

\author{I. Petkovi\'c}
\affiliation{Laboratoire de Physique des Solides, UMR 8502, Universit\'{e} Paris-Sud,
B\^{a}t. 510, 91405 Orsay Cedex, France.}


\begin{abstract}
Hybrid Superconductor/Ferromagnet structures have been investigated recently to
address the interplay between ferromagnetism and superconductivity. They also open
up new routes for the investigation of out of equilibrium superconductivity. Here,
we show how it is possible for out of equilibrium excitations produced
in a superconducting thin film (S) to be localized in a ferromagnetic trap (F).
Specifically, a ferromagnetic nano-volume in good contact with S represents a
potential well for the quasiparticles (QPs) at the gap edge. As the
superconducting proximity effect is highly suppressed in F, QPs get efficiently
trapped and they share their energy with the free electrons in the trap. The
electronic temperature $T_e$ in the trap can be increased by up to $60\%$ from the
bath temperature at $320\,$mK as measured by tunneling spectroscopy using a second
junction.
\end{abstract}

\pacs{PACS numbers: 74.45.+c, 75.50.+r}

\maketitle

\section{Introduction}
\noindent Superconductor/Ferromagnet (S/F) hybrid nanostructures open up new
routes to study the interplay between superconductivity and ferromagnetism. The
main advantage is that unlike in bulk materials, in nanostructures despite the
fact that superconducting gap ($\Delta \sim \rm meV$) and ferromagnetic exchange energy
($E_{ex} \sim \rm eV$)  are not
comparable, superconductivity and ferromagnetism do not exclude one another. This
is because these two antagonistic ground states influence each other only near the
interface. Nonequilibrium superconductivity in such structures is a new topic.
After a superconductor absorbs energy, excess quasiparticles (QPs) with non-
thermal distribution are created. Their energy relaxation occurs via two different
mechanisms: electron-electron ($e$-$e$) and electron-phonon ($e$-$ph$)
interactions. The timescale for energy relaxation depends on the respective
scattering rates ( $\tau_{ee}^{-1}$ and  $\tau_{e-ph}^{-1}$), both of them rapidly
increasing with the excitation energy. \cite{ref1}  The Fermi-Dirac energy
distribution of QPs is reached on a timescale larger than  $\tau_{ee}$, while in
the limit of strong $e$-$ph$ relaxation, QP temperature becomes equal to the
lattice temperature. \cite{ref2}

\noindent To be more specific, when a superconductor absorbs energy $E_0$, within a few picoseconds it goes into a nonequilibrium state, consisting of broken Cooper pairs and high energy phonons with energies close to the cutoff Debye energy. \cite{ref3} In a
fast nanosecond cascade, energy is distributed over an increasing number of QPs
and phonons, decaying until it reaches the gap edge. Both excess QPs and phonons
have limited lifetimes. Quasiparticles may recombine into Cooper pairs, diffuse,
get scattered by impurities, or get trapped and frozen, while phonons can get
scattered or escape into the substrate. The characteristic timescale of QP
recombination into Cooper pairs goes from the ns until a few $\rm \mu s$ depending
on material parameters. \cite{ref1} If the superconductor $S_1$ is coupled to
another superconductor $S_2$ via a tunnel junction, excess carriers can tunnel
through the barrier. At low temperatures ($T<<T_C$, where $T_C$ is the
superconducting critical temperature), recombination time is longer than the
tunneling time, \cite{ref2,ref4} therefore majority of QPs can tunnel, creating
current through a tunnel junction. As the magnitude of current from a detector
tunnel junction is inversely proportional to the volume of the source
superconductor $S_1$, \cite{ref5} in order to enhance the readout signal, Booth
proposed the idea of QP trapping in a small volume layer that can be either a
superconductor with a smaller gap than that of $S_1$ or a normal metal.
\cite{ref6} We show a typical quasiparticle trapping device on Fig.\ref{fig1}.
The current is injected into the device via a tunnel junction from superconductor
$S_{inj}$, and detected through another tunnel junction with superconducting
electrode $S_2$.

\noindent We have considered a ferromagnetic trap ($F_{tr}$) and investigated the
energy confinement produced by downconversion of QPs coming from $S_{1}$ into
$F_{tr}$. \textit{As no gap is induced in $F_{tr}$ by proximity effect, a
ferromagnetic trap allows QPs to relax their energy in a volume much smaller than
that of a normal trap, increasing the QP accumulation and hence the local energy
density.} We have fabricated three terminal Josephson tunnel devices of different
sizes and geometries. QPs are injected into $S_{1}$ from a tunnel junction, they
relax at the gap edge and then they are trapped in $F_{tr}$ where through $e$-$e$
interaction they locally raise the electron temperature $T_{e}$. The temperature of these nonequilibrium carriers in $F_{tr}$ is probed by tunneling spectroscopy, measuring the current-voltage characteristics of a second tunnel junction connecting
$F_{tr}$ layer and $S_2$. We show that in mesoscopic devices, ferromagnetic
trapping is efficient in raising the local electronic temperature from the
bath temperature of $320\,$mK up to $580\,$mK.

\noindent The paper is organized as follows. In section II, we describe the
principle of quasiparticle trapping. In section III we present the principle of
operation of three terminal tunnel junction devices using a ferromagnetic trapping
layer. In section IV we describe our three terminal macroscopic device consisting
of two stacked tunnel junctions and present its characterization in terms of
conductance variation with QP injection. In section V we present our planar
mesoscopic three terminal device, consisting of two tunnel junctions connected
with a sub-micrometric S/F bilayer. We also measure conductance as a function of
steady-state current injection. In section VI we discuss the results and trap
heating in terms of a simple non equilibrium model.

\section{Principle of Quasiparticle Trapping}

\begin{figure}[h]
\centerline{\hbox{
 \psfig{figure=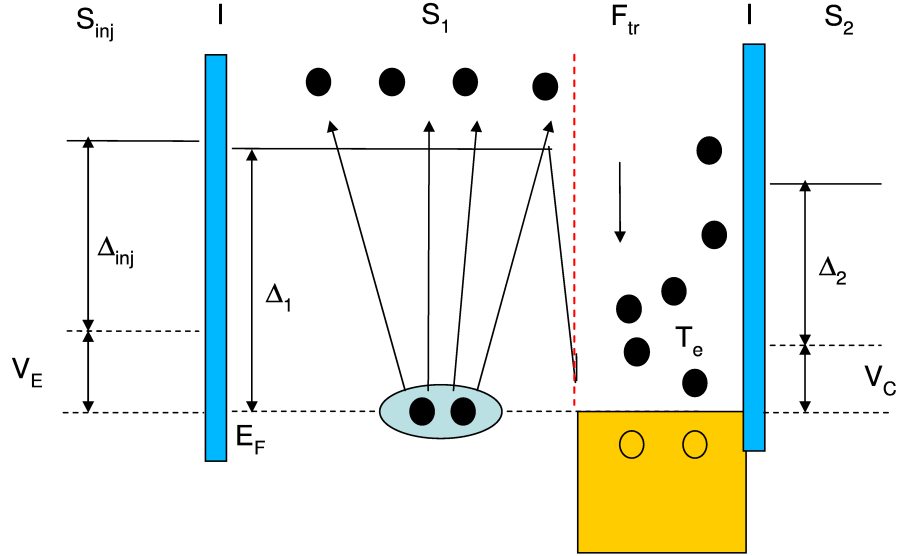,height=75mm,width=120mm,angle=0}}}
 \caption{ (Color online) Principle of quasiparticle trapping. Quasiparticles coming from the
superconducting reservoir $S_1$ with the energy comparable to the gap fall into the
potential well $F_{tr}$, and interact with the electrons, heating them. A tunnel
barrier $I$ in direct contact with the trap electrode allows the readout of the
nonequilibrium current signal. The measurement principle is also shown.
We are able to bias the two junctions in different configurations. \label{fig1}}
\end{figure}

\noindent As first proposed by Booth, \cite{ref6} trapping of excitations in a
small junction has been considered as an alternative to fabrication of large area
detectors, whose high capacitance limits both the frequency response and the
energy resolution. Excess QPs created by either Cooper pair breaking or injection
(see Fig.\ref{fig1}) diffuse from $S_1$ to the trap, where they relax their
energy and are confined. From this intuitive physical picture, it is clear that the trap layer has two crucial functions: it prevents back diffusion of QPs, and with its smaller volume it increases the magnitude of the readout signal, since the density of out of equilibrium quasiparticles $N^{QP}$, for a fixed energy, is proportional to the
inverse of the trap volume $N^{QP}=I_{inj} \tau_{tr}/e V_{tr}$, where $I_{inj}$ is
the injected current, $V_{tr}$ the trap volume and  $\tau_{tr}$ the time spent in
the trap before relaxation. However, if the volume is too small, phase
coherence is preserved in the trap resulting in a superconducting proximity
effect. The induced gap in the density of states in the trap reduces the number of
QPs available to relax the energy, and hence heating.

\noindent The choice of materials for the Superconductor/Trap ($S_1/T$) bilayer
is, therefore, crucial for correct operation. Firstly, $S_{1}$ will be chosen for
long QP lifetimes and high QP diffusion velocities to ensure that QPs reach the
trap before recombination. Secondly, it is important that interfaces between
$S_{1}$ and $T$ are clean, to facilitate the diffusion. Thirdly and most
importantly, we need a small superconducting proximity effect in the trap layer.
We should make a trap layer sufficiently thin in order to get a high density of nonequilibrium QPs, but not too thin as it should remain in
the normal state in contact with $S_{1}$ through a highly transparent interface.
Despite considerable efforts in several laboratories, the right compromise between
these constraints is still under investigation.

\noindent N.E. Booth et al. \cite{ref7} and Ullom et al. \cite{ref8,ref9} have
extensively studied normal trap efficiency, defined as the ratio between detected
and injected power, in a system consisting of an Al superconducting reservoir
$S_{1}$, and Ag as the normal metal trap. They used a three terminal device
following the diagram shown in Fig.\ref{fig1}. Tunneling
through a detector junction determines the dwell time and the speed of the
detector, both of the order of $\rm \mu s$. Ullom et al. described the QP
diffusion in terms of a renormalized, energy dependent diffusion coefficient $D_E=
(v(E)/v_F)D$, where $v(E)=v_F[1-(\Delta/E)^2]^{1/2}$ is the QP group velocity that
drops to zero at the gap edge, $\Delta$  is the superconducting gap, $v_F$ the
Fermi velocity, $E$ the energy measured from the Fermi level and $D$ is the bulk
diffusion coefficient. They have measured at $100\,$mK up to $17 \%$ efficiency. Because of the large device size ($\sim100\, \rm \mu m$), QP losses were mainly due to
recombination. In another series of experiments with excitations produced in In
crystals by a low power pulsed laser, Goldie et al. \cite{ref5} studied trapping
into Al and Cu thin films ($250$ to $500\,$ nm) deposited on In always finding a minigap induced in the Cu by the proximity effect. Pepe et al. \cite{ref10} reported nonequilibrium measurements under steady state current injection in a stacked double
superconducting tunnel junction device with middle electrode consisting of an
Nb/Al bilayer where the Nb is the superconductor and the Al the trap. They found
that the $20$ nm thick Al trap layer was strongly proximised by a $40$ nm thick
Nb film. They also found that the proximity effect is reduced as a function of the
injection current, without any appreciable modification of the interface
transparency.  Zehnder \cite{ref11} has studied how external energy is stored for
different junction materials. He found losses mainly due to the presence of the
sub-gap phonons in the trapping layer. Efficiency scales inversely with the mean free path of the material. He also pointed out that for large area ($50\times50\, \rm \mu m^2$) Nb/Al structures the density of states is spatially dependent. As a consequence, the trapping process becomes local, and hence both QP lifetime and trapping rate.

\noindent In the present work we propose a new direction, the reduction of the
trap volume using a ferromagnet $F_{tr}$. The superconducting proximity effect
extends in the ferromagnet  over a coherence length given by  $\xi_F = (\hbar D/E_{ex} )^{\frac{1}{2}}$ that varies between $0.1$ and $10\,$nm and is almost $100$ times smaller than the coherence length of a normal metal with the same diffusion constant, $D$. This extremely reduced proximity effect provides a more efficient energy localization. So, a hybrid, mesoscopic S/F heterostructure seems to be a unique candidate to obtain the smallest, not proximised trap.

\section{Three terminal device operation}

\noindent Here we describe the working principle of our realization of the
ferromagnetic quasiparticle trapping transistor (FQTT) based on the description
proposed by Booth et al. \cite{ref13} The ferromagnet is a $\rm Pd_{0.9}Ni_{0.1}$
alloy. \cite{ref14} We choose this alloy because: $1$. the small ferromagnetic coherence length ($\sim 10\,$nm); $2$.  Pd has an almost filled $d$ band, and the large
density of states at the Fermi level provides more conduction electrons available
in the trap; $3$. electronic specific heat is high ($\sim 3$ orders of magnitude
higher than pure Copper, for example), \cite{ref15} which means that energy is
"stored" in the electron gas, with substantially longer $e$-$ph$ relaxation time. It is worthwhile to point out that long $e$-$ph$ relaxation time leads to a small bandwidth.

\noindent The principle of operation of our devices can be obtained considering a
simple static balance of power flow into the $F_{tr}$ layer.
To do this, we borrow the common semiconductor transistor terminology. We call
$S_{1}$/$F_{tr}$ bilayer the \textit{base}, $S_{inj}$ the \textit{emitter}, and
$S_2$ the \textit{collector}. Current $I_{inj}$ is passed through the first
junction emitting $I_{inj}/e$ quasiparticles. The net power injected into $S_{1}$
is $P_{inj}=(I_{inj}/e)\,\varepsilon_{inj}$, where $\varepsilon_{inj}$ is the
average injected QP energy, which depends on the voltage $V$ across the tunnel
junction. A part of the incoming power is lost via thermal interaction with the phonon lattice, $P_{th}$. It can be calculated using the so called $2\,$-T nonequilibrium model, \cite{ref16} in which different temperatures $T_e$ and $T_{ph}$ and specific heats $C_e$ and $C_{ph}$ are assigned to the electron and phonon systems, which are
considered to be decoupled at low temperatures. The energy transfer rate from
electrons to phonons in the $S_{1}$/$F_{tr}$ bilayer is given by $P_{th}=\Sigma \,
V_{tr}\,({T_e}^5-{T_{ph}}^5)$, where $\Sigma$   is a material-related  parameter
(varying in the range of $1$-$5\,\rm nW/ \mu m^{-3}K^{-5})$ and $V_{tr}$ is the
volume of the trap. Here we explicitly assume that the phonon system is at the
external bath temperature. This hypothesis, as we shall see later, depends on the
details of the film geometry and coupling to the substrate, via the Kapitza
interface conductance.\cite{ref17} Injected power is also evacuated via the
detector tunnel junction, by cooling through tunneling. This happens when a NIS
junction is biased at a voltage $V$ lower than the superconducting gap, so that
only the hot electrons from the tail of the Fermi distribution can tunnel from the
normal to the superconducting electrode. Self cooling, by removing high energy
electrons, causes a progressive transfer of energy out of the normal film.\cite{ref18} The cooling power of the junction when polarized at the gap
edge is given by $P_{Cool} \approx (0.5 \, {\Delta_2}^2\,/e^2 R_T )\,(k_B T/
\Delta_2)^\frac{3}{2}$, \cite{ref19} where $R_T$ is the normal state resistance of
the junction. In the experiments described below the cooling power can be neglected and hence as we shall see below all the injected energy is lost by
phonon interactions:

                      $$(I_{inj} /e)\, \varepsilon_{inj} - \Sigma \, V_{tr}\,
({T_e}^5-{T_{ph}}^5) \approx 0,$$

\noindent which yields:
\begin{equation}
\label{eq2}
T_e=T_{ph} \left[\,1+\frac{( I_{inj} /e)\, \varepsilon_{inj}}{ \,\Sigma\, V_{tr}\,
{T_{ph}}^5}\right]^\frac{1}{5}.
\end{equation}

\noindent If high energy QPs are injected with $\varepsilon_{inj}\gg \Delta_1 $ ),
the most efficient mechanism
for energy redistribution within the electron subsystem becomes the emission of
Debye phonons. The mean free path of these phonons is very small, and they
efficiently excite additional electrons and break Cooper pairs in the $S_{1}$
reservoir.  Nonequilibrium vibrations isotropically propagate through the whole
system, mixing and coupling the electronic and phononic subsystems. As a result,
delocalized heating takes place and energy confinement is no longer
possible. Therefore it is important to work at energies near the gap edge in order
to avoid phonon losses (which multiply with the increase of the injection energy)
and transfer the injected energy solely to the electronic system.

\section{Stack junction device}

\noindent We have fabricated two kinds of three terminal structures with different
geometries and sizes. In both we chose Al as superconductor $S_{1}$ for its long
recombination time\cite{ref20}.
\cite{ref1}

\begin{figure}[h]
\centerline{\hbox{
 \psfig{figure=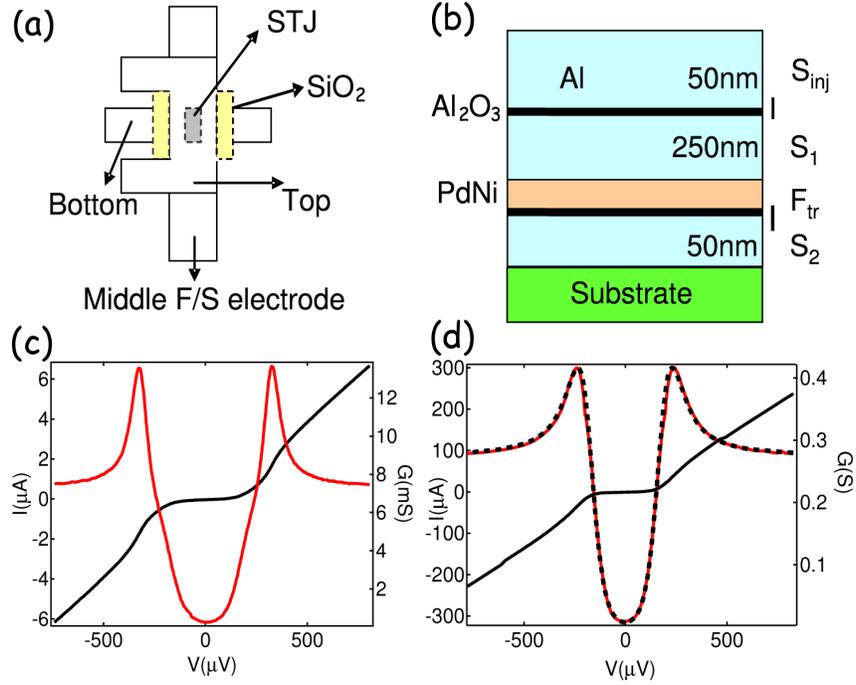,height=94mm,width=115mm,angle=0}}}
 \caption{Sketch of the top view  (a) and the cross-section (b) of the stacked
device. The junction area is $0.7\times0.7 \, \rm mm^2$ and the distance between junctions on
the same sample is about $1\, \rm mm$.
Experimental I-V curves (black) and conductance spectra (red curves) of the top
 (c) SIS and bottom (d) SISF junction of the stacked
device. Normal resistances are, respectively, $140\,\Omega$ and $4\,\Omega$ . (d) the BCS fit (dashed line) of the conductance spectrum at $T=0.32\,$ K and gap value of $0.195\, \rm meV$ is also shown.
\label{fig2}}
\end{figure}

\noindent The first macroscopic device that we shall present consists of two
stacked junctions (area of $0.7\times 0.7\,\rm mm^2$). We put electric contacts so
that we can bias each junction independently. The stacked structures used for the
experiment described in the following are sketched in Fig. \ref{fig2} in a top (a)
and cross-section (b) view. Four stacked junctions are fabricated simultaneously on a
single Si wafer, with common bottom electrode.  All layers were fabricated in situ
using thin film evaporation, in an ultra high vacuum system with a base pressure
of $10^{-9}\,\,\rm mbar$. The fabrication procedure is as follows. On a Si substrate
we evaporated through a metallic mask the first $50\,$nm-thick Al layer. This
layer will be the bottom electrode $S_2$ of the device, common to all the
junctions of the sample. Then we oxidize the Al layer by $\rm O_2$ plasma at about
$8\times 10^{-2}\,\rm mbar$ pressure. The junction area is defined by evaporating
$50\,$nm-thick orthogonal striplines of SiO in order to define a square of $0.7
\times 0.7\,\rm mm^2$. Then, a layer of PdNi ($10 \%$ Ni) is evaporated as
ferromagnetic trap $F_{tr}$. We have varied the thickness ($5$, $8$, $10\,\,$nm) of
the $\rm Pd_{0.9}Ni_{0.1}$ (PdNi) layer. The trap volume is $V_{tr} = 2500\, \rm \mu
m^3$ for $5\,$nm of PdNi. Then we evaporated $250\,$nm of Al as superconducting
layer $S_{1}$. We repeated the oxidation and SiO deposition in order to make
a second tunnel barrier, and finally evaporated $50\,$nm of Al as layer $S_{inj}$.

\noindent The junctions quality was systematically checked by measuring the
current-voltage characteristics and the tunnel conductance with the standard ac
modulation technique. All the measurements were carried out at $320\,$mK in a
$\rm ^3He$ cryostat. In Fig.\ref{fig2} (c) and (d) are shown two typical I-V curves and tunnel conductance spectra of an $\rm Al/Al_2 O_3/Al$ (SIS) junction and an $\rm
Al/Al_2 O_3/PdNi$-$\rm Al$ (SIF) junction, respectively. Quasiparticle current is suppressed below $360\, \rm \mu V$, i.e. twice the Al gap for the upper SIS junction, while for the bottom SIF junction we measure a voltage gap of $195\, \rm \mu V$ showing no proximity effect in F. The tunneling spectrum of the bottom junction is well
fitted by the conventional BCS density of states (dashed line), instead the upper junction shows that the Al on the PdNi has not fully recovered its BCS density of states and states in the gap are partially occupied. Normal resistances
are $140 \, \Omega$  for SIS and $4 \, \Omega$  for SIF junction. The critical
temperature of the PdNi/Al bilayer is $1.2\,\, \rm K$.

\begin{figure}[h]
\centerline{\hbox{
 \psfig{figure=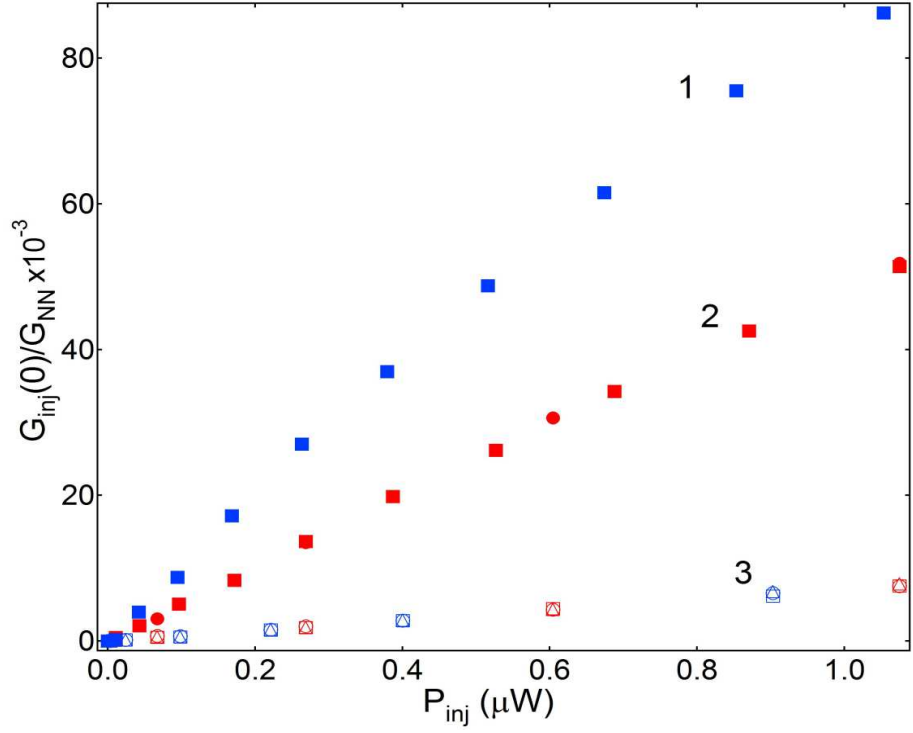,height=100mm,width=130mm,angle=0}}}
 \caption{Tunneling conductance variation at zero bias (normalized to the normal
state conductance) as a function of the injected power. Marker groups (1) and (2) refer
to the injection and detection in the same stack. Markers (1) corresponds to the
injection from the SIF and detection at the SIS and markers (2) vice versa. Markers
(3) refer to the injection and detection between junctions of different
stacks.\label{fig3}}
\end{figure}

\noindent We measured the dynamical conductances of the bottom junction
under injection of current between $0$ and $600\, \rm \mu A$ from the top junction.
We did not observe any enhancement in the detection spectra at energies close to the gap, while at energies $50$ or even $100$ times larger, zero bias conductance increases linearly in power (variations of about $10 \%$ of the normal state conductance are measured) due to hot electrons. We have also reversed the
roles of the junctions. The injected power is of the order of  $\rm \mu W$
in both configurations. As shown in Fig. \ref{fig3}, marker groups (1) and (2) refer to
the injection and detection in the same stack. Markers (1) corresponds to injection
from the SIF and detection at the SIS and markers (2) to the inverse. As $E>\Delta $, phonon emission drives the relaxation process, we measure
twice higher signal when injecting from the SIF (1) rather than the SIS (2)
junction because of the double convolution by the QP density of states of the two
superconducting electrodes.

\noindent Markers (3) refer to injection and detection between junctions of
different stacks. All marker groups (3) are practically superimposed, regardless of the
direction of injection. No differences are observed in the detector as a function
of the distance from the injector. Therefore even though energy is not localized
in $F_{tr}$, the lack of accumulation is not due to the out diffusion of the QPs
away from the detector junction area into the surroundings. Indeed, from the recombination time  $\tau_r =100\, \rm \mu s$ \cite{ref1} which sets the upper limit
for QP diffusion, we found diffusion length $L_r\sim 10\,\rm \mu m$, which is small
compared to the dimensions of our device. Moreover, we did not register any difference between the injection from the top/bottom junction of the stack and detection with the bottom junction of the neighboring stack. This implies, as the bottom electrode is geometrically common to all the junctions of the sample, that the tiny raise in the tunneling conductance (i.e. markers (3)) is mainly due to a small rise in the substrate temperature (we also did not measure any thermal gradient between neighboring junctions). Let us consider the power balance and the dynamics of excitations in the PdNi trap. We calculate specific heat of electron and phonon subsystems: $C \rm (mJ/molá K) = $ $\gamma T + \beta T^3$, where the linear contribution is due to electrons and cubic to phonons. From the molar volume of the trap ($8.64\, \rm cm^3/mol$) we have: $\gamma_M  = 1.38\cdot 10^{-15}\, \rm Já \mu m^{-3} K^{-2}$ and $\beta_M = 1.7 \cdot 10^{-17}\, \rm J \mu m^{-3} K^{-4}$ and hence: $C_e = 1.0 \cdot 10^{-12}\,\rm J/K$ while $C_{ph}  = 1.2 \cdot 10^{-15}\,\rm J/K$ at $300\,$mK. We found that the electron-phonon conductance ($dP_{ph}/dT_e \equiv G_{e-ph} = 5 \Sigma (A d) {T_e}^4$) at $T_e = 300\,$mK is $0.2 \,\rm \mu W/K$ and the electron-phonon scattering time ($\tau_{e-ph} = C_e / G_{e-ph}$) is $5.1 \,\rm \mu s$. Once thermal excitation is transferred to the lattice, energy can flow by thermal conductance to the substrate. This transfer is limited by thermal boundary resistance, the Kapitza resistance, $R_K$. Typically, $R_K á T^3 = 5 \cdot 10^{-4}\, \rm K^4 \mu m^2 W^{-1}$. Thus the effective
conductance will be: $$G_K = \frac{\rm Area}{R_K} = \frac{T^3}{5 \times 10^{-4}} \left[\frac{\rm Area}{\rm m^2}\right]\,\rm W/K. $$

\noindent There are two boundaries to consider: PdNi to Al bottom electrode, and
bottom electrode to substrate. We neglect the boundary between the surface oxide
on the substrate and the single crystal Si wafer. So, taking as area a larger
value of about $1\, \rm mm^2$ and considering the series of two interfaces, we
obtain $G_K = 27\, \rm \mu W/K$, which implies $G_K$ about a factor $100$ larger
than $G_{e-ph}$. The phonon temperature is uniform on the sample. This implies
that our substrate has very good coupling to the $S_{1}$/$F_{tr}$ bilayer.

\noindent We now discuss why the ferromagnetic trap is not working properly. The
trap works at the gap edge and, in our case, at that energy the flow of injected
QPs is too small. In fact, the density of out of equilibrium QPs in the trap must be higher than that in the thermal equilibrium, $N^{QP}>{N_0}^{QP}$, for the trap to be effective. This condition imposes a lower bound on the lifetime of the
out of equilibrium excitations in the trap $\tau_{tr}$ as $N^{QP}=I_{inj}
\tau_{tr} /e V_{tr}$ and ${N_0}^{QP}=n( \varepsilon_f ) k_B T$ , where $I_{inj}$
is the injected current (typically of about $1\, \rm \mu A$ at the gap edge with our
resistance values), $V_{tr}$ is the trap volume, and $n(\varepsilon_f)$ is the
density of electron states at the Fermi level. We estimated that $\tau_{tr}$ of
about $300\, \rm \mu s$ is required to accumulate excitations with a number higher
than ${N_0}^{QP}$. This time is much larger than any characteristic timescale and
in particular $\tau_{e-ph}$, thus excitations out of equilibrium created by
injection are practically lost via phonons. The trap is inefficient
and the conductance at zero bias independent of injection at energy comparable
with the Al gap edge.

\noindent The first obvious solution is to increase the injected current density.

\section{Mesoscopic planar device}

\begin{figure}[h]
\centerline{\hbox{
\psfig{figure=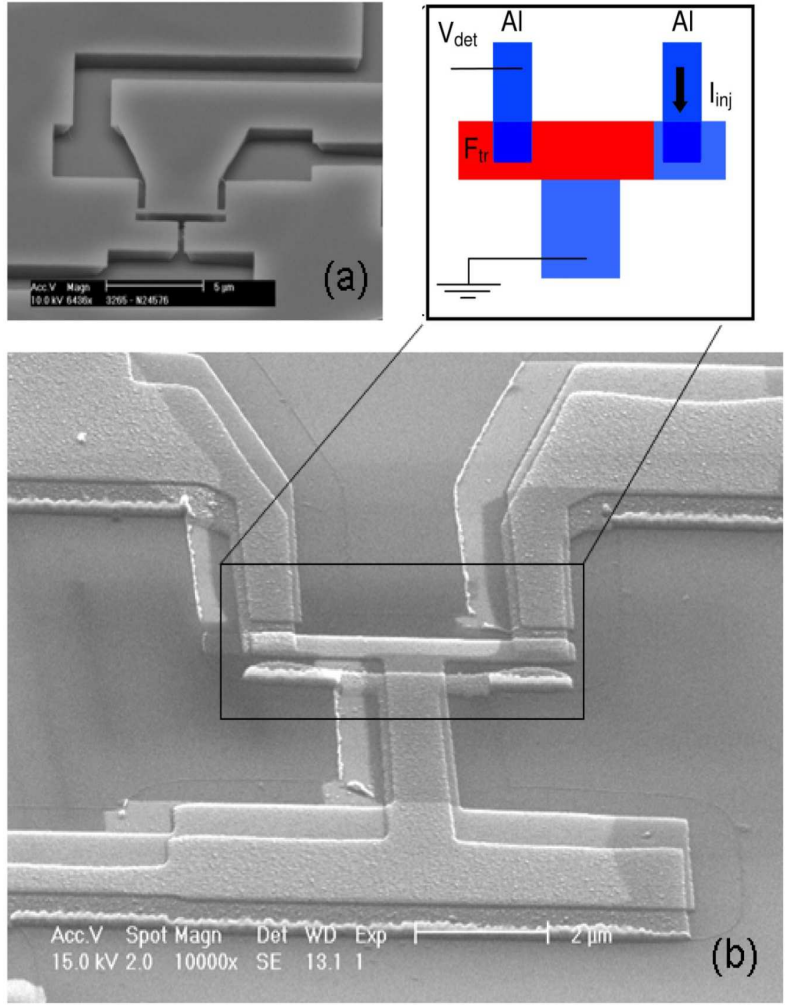,height=140mm,width=110mm,angle=0}}}
 \caption{(a) SEM image of the trilayer mask. (b) SEM image of the planar device. The $\rm Al/Al_2O_3/Al$ (SIS) junction on the right and $\rm Al/Al_2O_3/Al/PdNi$ (SISF) junction on the left are shown in the inset. They are connected by a $1 \rm \,\mu m$ wide Al/PdNi bilayer. (c) The schematics of the device. DC currents are injected from the right SIS and detected through the SISF junction.\label{fig4}}
\end{figure}

\noindent We fabricate a new type of mesoscopic planar device to get smaller
volumes and higher trap efficiency. This type of device is represented in the
photo of Fig. \ref{fig4}(b). Two sub-micrometric tunnel junctions are separated by
middle, double layered Al-PdNi common electrode, the trap.

\noindent The fabrication is as follows. We used a trilayer suspended mask
consisting of $600\,$nm of polymer polyether sulphone (PES), $60\,$nm $\rm Si_3
N_4$ layer, and $400\,$nm PMMA layer. The mask is patterned by electron beam
lithography and the pattern transferred to the $\rm Si_3N_4$ layer by $\rm SF_6$
plasma reactive ion etching at a pressure of $10\,$mbar for about $60$ seconds.
Then the PES is etched with $\rm O_2$ plasma at a pressure of $300\,$mbar for $10$
minutes, which gives an undercut of about $500\,$nm. The SEM photo of the mask is
shown on Fig. \ref{fig4}(a). Our samples were fabricated by angle evaporation in an
ultrahigh vacuum system with a base pressure of about $10^{-9}\rm mbar$. First
$50\,$nm of Al were evaporated. The Al was oxidized in pure $\rm O_2$ flux at a
pressure of $5\times10^{-2}\rm mbar$. Then a $50\,$nm-thick Al layer was
evaporated, followed by a $5\,$nm-thick PdNi layer. Four devices are fabricated on
the same wafer.
The schematic of the device is shown in Fig. \ref{fig4}(c).  The first, right-side $300\times 500\rm nm^2$ $\rm Al/Al_2O_3/Al$ junction is separated by another junction of the same area by a $4\,\rm \mu m$ long Al layer which is partially ($\sim 3\, \rm \mu m$) recovered by a
thinner ferromagnetic PdNi layer. The magnetic layer is used to suppress
superconductivity by inverse proximity effect and it defines the trap. Trap volume
is $V_{tr} = 0.15\, \rm \mu m^3$ i.e. about $100$ times smaller than the trap used
in the stacks.

\noindent Note that the F layer plays a slightly different role in this device,
since the QPs coming from the Al superconducting electrode tunnel into the Al
and then diffuse lateraly into the PdNi/Al bilayer made up of the same Al.
As the reservoir and the trap are made of the same material, matching of the Fermi velocities is optimal and so is the QP transmission. Furthermore, in this planar structure we deal with lateral trapping, which allows to avoid back tunneling of QPs, responsible for reduced cooling power. \cite{ref21}

\begin{figure}[h]
\centerline{\hbox{
 \psfig{figure=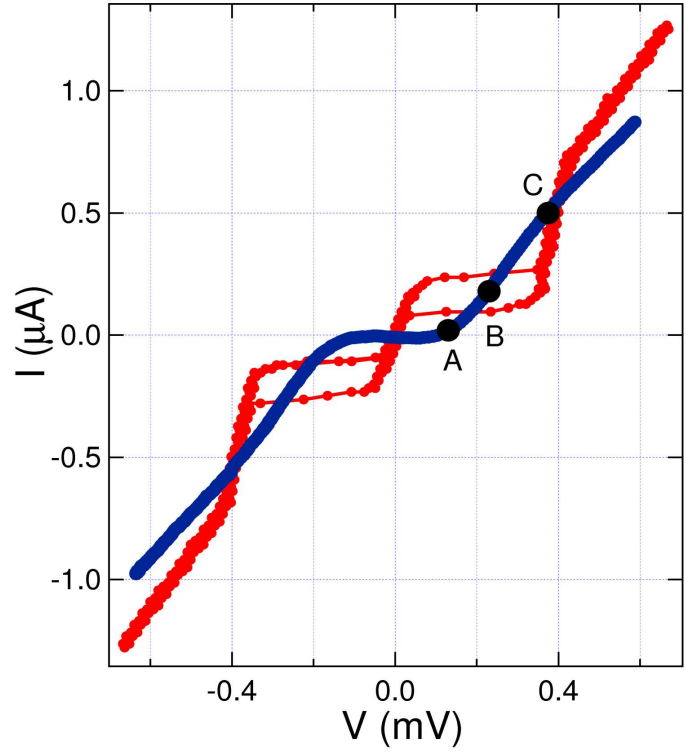,height=100mm,width=90mm,angle=0}}}
 \caption{I-V curves of the $\rm Al/Al_2O_3/Al$ (SIS) junction (red) and the $\rm Al/Al_2O_3/Al/PdNi$ (SISF)
junction (blue). The SIS junction shows a Josephson critical current of 200 nA et a retrapping current of 85 nA. \label{fig5}}
\end{figure}

\begin{figure}[h]
\centerline{\hbox{
 \psfig{figure=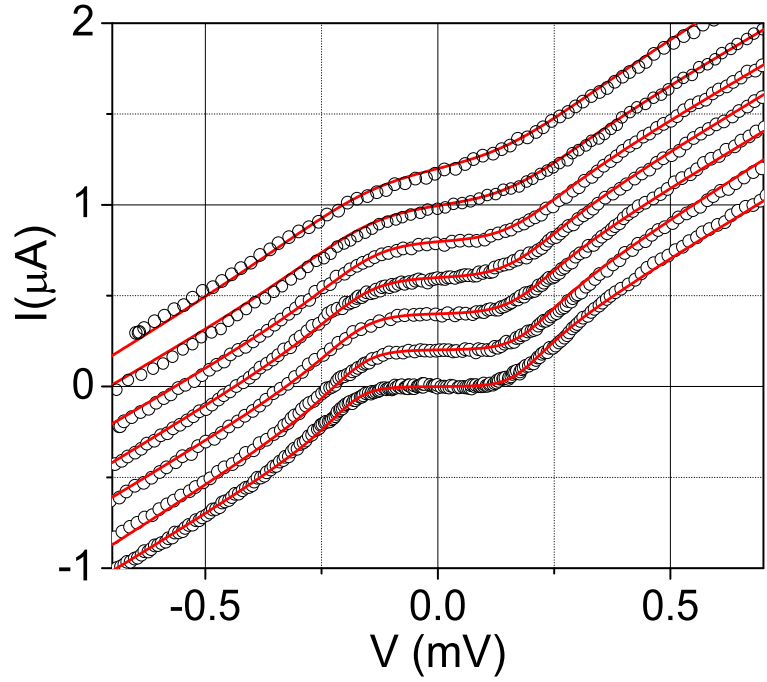,height=90mm,width=100mm,angle=0}}}
 \caption{Experimental I-V curves (dots) and BCS fits (red lines) of the detector
junction under a DC current injection between $0.3\, \rm \mu A$ and $8 \, \rm \mu
A$.\label{fig6}}
\end{figure}

\noindent In Fig. \ref{fig5}, the I-V curves of the $\rm Al/ Al_2O_3/Al$ (SIS) and
the $\rm Al/Al_2O_3/Al/PdNi$ (SI/SF) junctions are shown. The normal state
resistances are about $750\, \Omega$  and $500\, \Omega$ respectively. Junction
resistances higher than those measured in the stacks result in reduced injected
power (now ranging from $50\,$pW to $1\,$nW). The gap voltages are $400\,\rm \mu V$
and $200\,\rm \mu V$, as expected for inverse proximity effect. Moreover, the I-V
curve of the SI/SF junction is well fitted by integrating the BCS density of
states with an Al energy gap of $190\,\rm \mu eV$ and a base temperature of
$320\,$mK (see Fig. \ref{fig6}). The SIS junction is hysteretic with critical
current of $200\,$nA and retrapping current of $85\,$nA. The finite resistance at
zero bias is due to the bilayer trap resistance in series with the junction. The
principle of the experiment is the same as that described above. Out of equilibrium
excitations are collected in the SF trap rising the electron temperature $T_e$, which is measured by tunneling spectroscopy. Different conductance spectra were measured while increasing the DC injection current through the SIS junction from $0.3$ to $8\,\rm \mu A$, as shown in Fig. \ref{fig6}. We obtain $T_e$ from the BCS fits also shown in Fig. \ref{fig6} as red curves. The electron temperature $T_e$ as a function of the injected current is plotted in Fig. \ref{fig7}, top panel, markers. The temperature rises rapidly up to $0.53\,$K, and then reaches $0.98\,$K more slowly, resulting in two different slopes. Unlike stack junctions, the temperature in the planar trap increases
significantly for the injection of QPs with energy comparable with the
superconducting gap energy. Furthermore, the crossover between these two regimes
is set by a voltage bias corresponding to the gap edge as expected because
trapping, and hence energy localization, are more efficient below the gap edge. As
injection energy increases ($\sim 40$ times more for $I_{inj}=8\,\rm \mu A$), out of
equilibrium phonons are emitted before the QPs reach the trap. High energy
relaxation occurs mainly by phonons and phonon losses reduce the energy localized in the trap, resulting in a smaller temperature increase.

\section{Trap heating}

\noindent For $E<\Delta_{inj}$ quasiparticle energy relaxation in the trap is possible because, firstly, at $0.3\,$K the trap size is larger than $L_{ee} = (Dá \tau_{ee} )^{1/2} = 3\,\rm \mu m$, secondly, the electron-phonon scattering time is much longer than $\tau_{ee}\sim 10\,$ns and thirdly, the trap is located at a distance of about a
micrometer from the SIS junction. This distance is much smaller than the
recombination length $L_r \sim 100\,\rm \mu m$ as estimated above.  An estimation of
the electron-phonon scattering time from the electron-phonon conductance $G_{e-ph}
=1.6\,\rm pW/K$ and the specific heat $C_e =6 \times 10^{-15}\,\rm J/K$ gives
$\tau_{e-ph}= 0.5\,\rm \mu s$. The effective surface trap area evaluated from the
SEM image is about $3\,\rm \mu m^2$ and it gives $G_K = 160\,\rm pW/K$ Kapitza
conductance from the phonon bath to the substrate. As $G_K >> G_{e-ph}$, the
phonons in the device are thermalized at the phonon base temperature. The bandwidth of the device is limited by the smallest between thermal tunneling and electron-phonon time. \cite{ref13} In Table I are summarized all the relevant physical parameters that determine QP trapping, for both stacks and mesoscopic devices investigated here.

\begin{figure}[h]
\centerline{\hbox{
 \psfig{figure=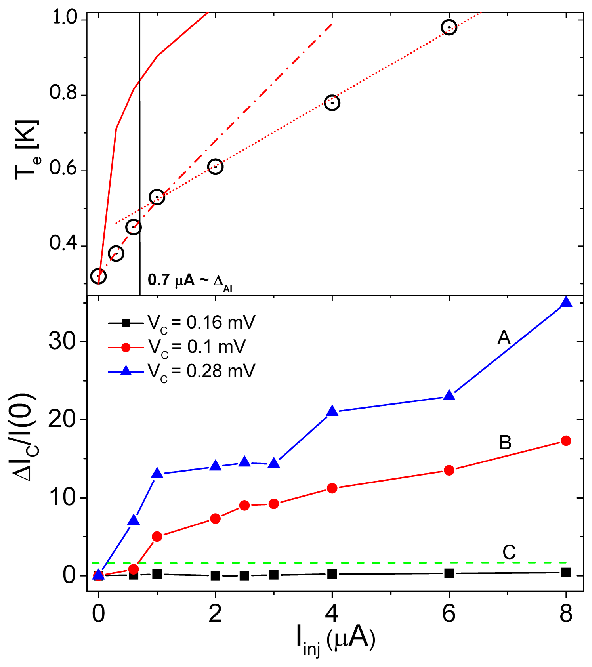,height=120mm,width=100mm,angle=0}}}
 \caption{(Top panel) Electronic temperature in the trap, as a function of the
injected current. Markers refer to the electronic temperature extracted by
the experimental conductance spectra via a BCS fit. The change in the slope occurs
approximately at $0.7 \,\rm \mu A$, corresponding to the Al gap. The full line
represents the expected increased temperature, calculated from the electron - phonon
coupling. The dotted-dashed line corresponds to the increase of the electronic
temperature due to the Al QP thermal conductivity. (Bottom panel) Extra-current (i.e.
the difference between the detected current with and without injection, normalized to
the respective value without injection:   $\Delta I_C/I_C(0)$) at three different
polarization values (A, B, C points are given in Fig. 5), as a function of the injected
current.\label{fig7}}
\end{figure}

\noindent In Fig. \ref{fig7} (top panel, full line) we report the temperature
increase as estimated from Eq. (\ref{eq2}) considering negligible cooling by
tunneling. (An estimate gives about $2\,$pW, a factor hundred smaller than the
injected power.) The actual temperature increase is lower than that obtained
considering only cooling by phonons (solid line). Instead, we found good agreement
by introducing a QP thermal loss (dotted-dashed line), $ \kappa á\Delta T$, where
$\kappa$   is the thermal conductivity of the Al leads whose normal state
resistance is $1.2\,\, \Omega$.

\noindent Now let us discuss the current gain. The excess current of the SI/FS
junction normalized to the value for zero injected current, $\Delta I_C/ I(0)$,
for three different voltage bias values, is shown in Fig. \ref{fig7}, bottom panel.
When biasing well above the gap voltage (curve C on Fig. \ref{fig7} corresponding
to point C on Fig. \ref{fig5}, $V_C= 0.28\,\rm mV$), no excess current is measured.
This range corresponds to the \textit{high energy tails} of the QP DOS and it is
almost independent of injection. When we approach the gap voltage we measure more
pronounced variations of the collected current. When we polarize the detector
junction at $V_C=0.1\,\rm mV$ (i.e. below the gap voltage, point A on
Fig. \ref{fig6}) we measure an increase of the tunnel current up to $15$ times. As
expected, it is at energies near the gap that the increase of current is maximum:
at  $V_C=0.18\,\rm mV$, point B on Fig. \ref{fig6}, the collected current is up to
$35$ times $I_C(0)$. It is important to note that the trapping mechanism not only
controls the number of out of equilibrium QPs, but also the direction of the
energy flow.

\noindent  If we reverse the roles of the junctions, injecting from the SI/SF
and detecting at the SIS, we do not register any increase of output current.
All the current values detected in the reversed configuration fall below
the dashed line in Fig. \ref{fig7}, low panel.  Therefore, the ferromagnetic trap
introduces a strong asymmetry in the device. Good isolation is very interesting for
applications. However the amplification parameter as defined by Booth, \cite{ref13} $\beta =I_C/I_{inj}$, where $I_C$ and $I_{inj}$ are, respectively, the collected and injected currents, normalized to the current gap values $I_{C0}=I_{inj0}= \Delta_{Al}/R_N$, is $0.24$ for detection at constant normalized voltage $V_C=0.8$. Higher amplification may be achieved at lower temperature and highly thermally isolated traps.

\section{Conclusion}

\noindent We have investigated a three terminal device consisting of two
superconducting tunnel junctions coupled via an S/F bilayer. The ferromagnetic
layer is used to localize and multiply QPs injected from the base junction. QP
excess current is measured through the second detecting tunnel junction. We
measured conductance spectra of the detecting junction as a function of the
injected power for two types of structures fabricated with the same S/F materials:
a stacked macroscopic and a planar mesoscopic device. We have found efficient
trapping in the mesoscopic device, where it is possible to increase the local
electronic temperature by up to $60\%$ of the base temperature. We have also found
that input is well isolated from the output as a consequence of using the
ferromagnetic trap, making the device attractive for transistor-like operation.
However the gain at $320\,$mK is $0.24$.

\begin{acknowledgments}
\noindent This work was partially done at the Quantum Physics Laboratory at the
ESPCI and it was supported by the ESPCI and the ESF through the "Pi-Shift"
project. It has also been partially supported by MIUR under the Project PRIN 2006
"Macroscopic Quantum Systems-Fundamental Aspects and Applications of
Unconventional Josephson Structures. N. Booth, J. Lesueur, B. Leridon, B. Reulet,
G. P. Pepe and A. Barone are gratefully acknowledged for several ideas and
discussions. We also thank S. Collin for technical assistance and J. Y. Prieur for
a critical reading of the manuscript.
\end{acknowledgments}

\begin{table}
\caption{The physical parameters which determine the QP dynamics in both traps are
summarized. First column refers to the stacked macroscopic devices, and the second
to the planar mesoscopic structures.\label{te}}
\begin{ruledtabular}
\begin{tabular}{c c c}
\textbf{Device}&\textbf{Macroscopic stacked}&\textbf{Mesoscopic planar}\\
Trap volume [$\rm cm^3$]&$10^{-9}$&$10^{-13}$\\
$G_{e-ph}\,[\rm nW/K]$&200&0.02\\
$\tau_{e-ph}\,[\rm \mu s]$&5&0.5\\
$G_C\,[\rm nW/K]$&2.2&0.01\\
$\tau_t\,[\rm \mu s]$&700&0.07\\
Injected power [\rm nW]&1000&1\\
\end{tabular}
\end{ruledtabular}
\end{table}

\clearpage

\clearpage

\newpage

\begin{thebibliography}{22}
\bibitem{ref1}
S. B. Kaplan, C. C. Chi, D. N. Langenberg, J. J. Chang, S. Jafarey, and
D. J. Scalapino, Phys. Rev. B {\bf 14}, 4854 (1976).

\bibitem{ref2}
R. Cristiano in \textit{Superconducting detectors} (Proceedings of
5th SceNet School, Spain, 2004).

\bibitem{ref3}
A. Zehnder, Phys. Rev. B {\bf 52}, 12858 (1995).

\bibitem{ref4}
\textit{Nonequilibrium Superconductivity, Phonons, and Kapitza
Boundaries}, edited by K. E. Gray (Plenum, New York, 1981).

\bibitem{ref5}
D. J. Goldie, N. E. Booth, C. Patel and G. L. Salmon, Phys. Rev. Lett.
{\bf 64}, 954 (1990).

\bibitem{ref6}
N. E. Booth, Appl. Phys. Lett. {\bf 50}, 293 (1987).

\bibitem{ref7}
J. N. Ullom, P. A. Fisher and M. Nahum, Nucl. Instr. and Meth. A {\bf
370}, 98 (1996).

\bibitem{ref8}
J. N. Ullom, P. A. Fisher and M. Nahum, Phys. Rev. B {\bf 58}, 8225
(1998).

\bibitem{ref9}
J. N. Ullom, P. A. Fisher and M. Nahum, Phys. Rev. B {\bf 61}, 14839
(2000).

\bibitem{ref10}
G. P. Pepe, G. Peluso, R. Scaldaferri, A. Barone, L. Parlato, R.
Latempa, and A. A. Golubov, Phys. Rev. B  {\bf 66}, 174509 (2002).

\bibitem{ref11}
A. Zehnder, Ph. Lerch, S. P. Zhao, T. Nussbaumer, E. C. Kirk and H. R.
Ott, Phys. Rev. B  {\bf 59}, 8875 (1999).

\bibitem{ref12}
S. E. Shafranjuk and I. P. Nevirkovets, IEEE Trans. on Appl. Sup.
{\bf 15}, 1051 (2005).

\bibitem{ref13}
N. E. Booth, P. A. Fisher, M. Nahum and J. N. Ullom, Supercond. Sci.
Technol. {\bf 12}, 538 (1999).

\bibitem{ref14}
A. W. Overhauser and A. I. Schindler, J. Appl. Phys. {\bf 28}, 544
(1957).

\bibitem{ref15}
G. Chouteau, R. Fourneaux, R. Tournier and P. Lederer,  Phys. Rev.
Lett. {\bf 21}, 1082 (1968).

\bibitem{ref16}
V. E. Gusev and O. B. Wrigth, Phys. Rev. B {\bf 57}, 2878 (1998).

\bibitem{ref17}
E. T. Swartz and R. O. Pohl, Rev. Mod. Phys. {\bf 61}, 605 (1989).

\bibitem{ref18}
F. Giazotto, T. T. Heikkila, A. Luukanen, A. M. Savin and J. P.
Pekola, cond-mat/0508093 v4.

\bibitem{ref19}
M. M. Leivo, J. P. Pekola, and D. V. Averin
Appl. Phys. Lett. {\bf 68} 1996 (1996)

\bibitem{ref20 }
K. Segall, C. Wilson, L. Li, L. Frunzio, S. Friedrich, M. C. Gaidis
and D. E. Prober,  Phys. Rev. B {\bf 70}, 214520 (2004).

\bibitem{ref21}
J. Jochum, C. Mears, S. Golwala, B. Sadoulet, J. P. Castle, M. F.
Cunningham, O. B. Drury, M. Frank, S. E. Labov, F. P. Lipschultz, H. Netel, and B.
Neuhauser, J. Appl. Phys. {\bf 83}, 3217 (1998).
\end{thebibliography}
\end{document}